\documentclass[intlimits,twoside,a4paper]{article}

\usepackage[cp1251]{inputenc}

\usepackage[eqsecnum]{cmpj3}

\usepackage{bm}

\issue{2022}{25}{4}{43701}
\doinumber{10.5488/CMP.25.43701}

\title[The antiferromagnetic phase transition in the layered Cu$_{0.15}$Fe$_{0.85}$PS$_3$ semiconductor]%
{The antiferromagnetic phase transition in the layered Cu$_{0.15}$Fe$_{0.85}$PS$_3$ semiconductor: experiment and DFT modelling%
}
\author[V. Pashchenko \emph{et al.}]{
        V. Pashchenko\orcid{0000-0001-8883-5595}\refaddr{label1}, 
        O. Bludov\orcid{0000-0002-2435-5039}\refaddr{label1}, 
        D. Baltrunas\orcid{0000-0001-9688-1719}\refaddr{label2}, 
        K. Mazeika\orcid{0000-0003-4792-8650}\refaddr{label2}, 
        S. Motria\refaddr{label3}, 
        K.~Glukhov\orcid{0000-0001-8795-0460}\refaddr{label3}\thanks{Corresponding author: \email{kglukhov@gmail.com}},  
        Yu. Vysochanskii\orcid{0000-0002-2501-1780}\refaddr{label3}.}
\addresses{
\addr{label1} B. Verkin Institute for Low Temperature Physics and Engineering of NAS of Ukraine, 47 Nauky Ave., 61103, Kharkiv, Ukraine
\addr{label2} Department of Nuclear Research Center for Physical Sciences and Technology, 231 Savanoriu ave., LT-02300, Vilnius, Lithuania
\addr{label3} Institute for Solid State Physics and Chemistry, Uzhhorod National University, 54 Voloshyn Str., 88000, Uzhhorod, Ukraine
}


\date{Received July 23, 2022, in final form October 15, 2022}

\begin{document}

\maketitle

\begin{abstract}
The experimental studies of the paramagnetic-antiferromagnetic phase transition through M\"{o}ssbauer spectroscopy and measurements of temperature and field dependencies of magnetic susceptibility in the layered Cu$_{0.15}$Fe$_{0.85}$PS$_3$ crystal are presented. The peculiar behavior of the magnetization --- field dependence at low-temperature region gives evidence of a weak ferromagnetism in the studied alloy. By the ab initio simulation of electronic and spin subsystems, in the framework of electron density functional theory, the peculiarities of spin ordering at low temperature as well as changes in interatomic interactions in the vicinity of the Cu substitutional atoms are analyzed. The calculated components of the electric field gradient tensor and asymmetry parameter for Fe ions are close to the ones found from M\"{o}ssbauer spectra values.
The Mulliken populations show that the main contribution to the ferromagnetic spin density is originated from $3d$-copper and $3p$-sulfur orbitals. The estimated total magnetic moment of the unit cell (8.543~emu/mol) is in reasonable agreement with the measured experimental value of $\sim9$~emu/mol.
%
%
%\keywords Up to six keywords (\href{https://physh.aps.org/browse}{Physics Subject Headings})
\printkeywords
%
%\pacs 76.80.+y, 61.18.Fs, 61.72.-y, 71.15.Mb, 75.25.+z, 75.50.Pp
\end{abstract}

\section{Introduction}
%\doclicenseThis

Two-dimensional (2D) van der Waals (vdW) materials offer possibilities to study novel physical properties and explore their potential applications in electronic, optical, and spintronic devices in the nanoscale~\cite{LL1}. The realization of magnetism in easily exfoliated layered crystals provides accessibility to control and manipulate magnetic properties at a single atomic layer level~\cite{LL2, LL3, LL4, LL5, LL6}. The presence of multiferroicity in such materials, when they exhibit two or more primary ferroic properties, is important for potential applications in the non-volatile storage devices controlled by an external electric field.

Recently, 2D ferroelectric polarization was found in CuInP$_2$S$_6$ several layers flakes and even in monolayers~\cite{LL7, LL8}. On the other hand, 2D antiferromagnetism is also demonstrated for CuCrP$_2$S$_6$ layers~\cite{LL9}. Furthermore, multiferroic material can be prepared by doping or modifying some monolayers, such as black phosphorus and graphene~\cite{LL10}. 2D materials with spontaneous ferromagnetism and ferroelectricity have rarely been reported. Recenrly, it was found that 2D CuCrP$_2$S$_6$ is multiferroic with magnetism and ferroelectricity stems from Cr and Cu cations, and the magnetoelectric coupling follows from the spin-orbit interaction~\cite{LL11}. Copper chromium thiophosphate CuCrP$_2$S$_6$ is an antiferromagnetic-antiferroelectric multiferroic involving collective ordering mechanisms of magnetic Cr$^{3+}$ ions and off-centered Cu$^+$ ions, respectively~\cite{LL12}. The indium compound CuInP$_2$S$_6$ belongs to the same C2/c space group as CuCrP$_2$S$_6$ at room temperature, but due to a specific second-order Jahn--Teller instability of Cu$^+$, it attains a ferrielectric structure with Cc symmetry below $T_c \approx 315$~K. The solid solutions CuCr$_{1-x}$In$_x$P$_2$S$_6$ reveal disordered dipolar glass phases, because of randomness and frustration, and quasimolecular magnetic properties~\cite{LL13}. Dynamic polar clustering occurs in these solid solutions and superposes structural glassiness to the ferrielectric long-range Cu$^+$ order at low temperatures.

Transition metal phosphorus trichalcogenides MPS$_3$ (M $=$ Mn, Fe, Co, Ni, $\dots$) have monoclinic crystal structure (space group of C2/m), in which the metal cations (M) are surrounded by an octahedral cage of (P$_2$S$_6$)$^{4-}$ bipyramids, and the neighboring metals have a 2D honeycomb lattice arrangement~\cite{LL5}. The crystal layers stack with vdW forces. Resulting from the competitions between the direct M--M exchange and indirect superexchange, mediated through S$^{2-}$ anions within each layer, as well as the interlayer exchange, determine antiferromagnetic (AFM) ordering temperature T$_N$ and its type~--- zigzag, Neel or stripy pattern~\cite{LL14}.

In 2D materials, magnetic anisotropy is also crucial in establishing a long-range correlation. For FePS$_3$ compound, the trigonal distortion combined with the spin-orbit coupling yields a large single-axis magnetic anisotropy~\cite{LL15}, and it can be described by the Ising model. In this compound, the long-range order is present in the direction perpendicular to the crystal layers, and FePS$_3$ has a zigzag type of antiferromagnetic ground state.

Incorporation of atomic defects and chemical substitutions in MPS$_3$ 2D materials could manipulate and control their magnetic properties~\cite{LL16, LL17, LL18}.  Distinct magnetic order, spin direction, and magnetic anisotropy, exotic phases and properties are expected to be revealed in solid solutions of these layered crystals. For example, spin glass behavior was found in Fe$_{1-x}$Mn$_x$PS$_3$~\cite{LL19, LL20} and in CuCr$_{1-x}$In$_x$P$_2$S$_6$~\cite{LL13}.

In this paper there are presented magnetic properties of FePS$_3$ as a function of temperature, field and dilution of the magnetic atoms by means of substitution of a non-magnetic species, in this case copper~--- we studied the magnetic properties of 2D vdW layered Cu$_{0.15}$Fe$_{1.85}$PS$_3$. The crystalline flakes with stoichiometry Cu$_{0.15}$Fe$_{1.85}$PS$_3$ were grown using the gas transport method~\cite{L1}. By dielectric, specific heat, and ultrasonic measurements the structural phase transition close to 109~K was found in these crystals~\cite{L1}. Obviously, similarly to FePS$_3$, it should be the antiferromagnetic phase transition. In this work, using the M\"{o}ssbauer spectroscopy and magnetic investigations, together with first-principles studies, the peculiarities of magnetic ordering in the Cu$_{0.15}$Fe$_{1.85}$PS$_3$ alloy are traced with the aim to search for a new layered multiferroic material for nanoscale devices.

\section{M\"{o}ssbauer data}

M\"{o}ssbauer spectra were measured using $^{57}$Co(Rh) source in the transmission geometry. The sample was composed of not grinded separate plates. The low-temperature spectra were obtained using the closed cycle He cryostat (Advanced Research Systems, Inc.). The doublet and Hamiltonian were applied to fit the M\"{o}ssbauer spectra using WinNormos Site software~\cite{L3}. Isomer shift is presented relative to $\alpha$-Fe at room temperature.

\begin{figure}[htb]
	\centerline{\includegraphics[width=0.55\textwidth]{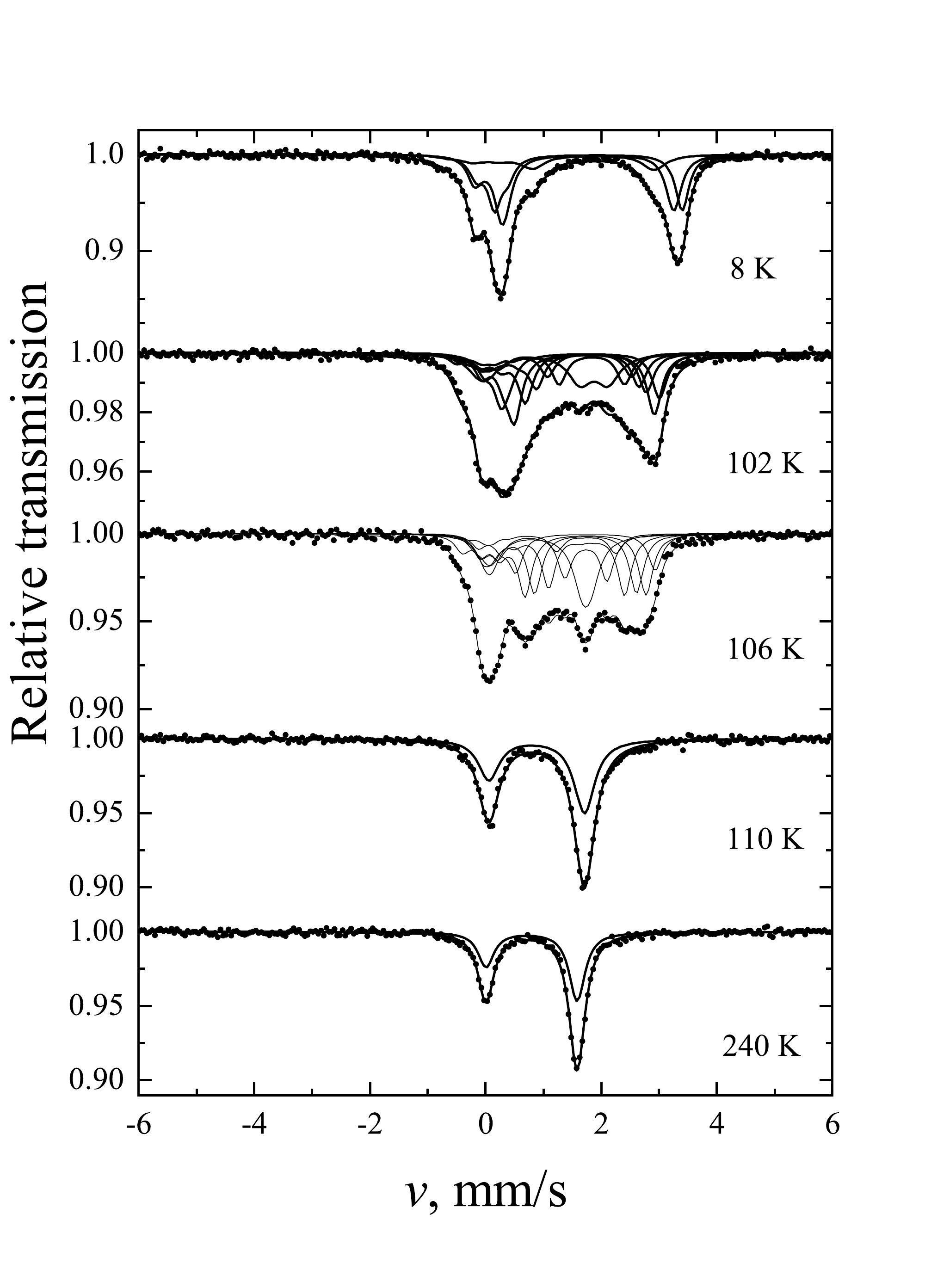}}
	\caption{M\"{o}ssbauer spectra of Cu$_{0.15}$Fe$_{0.85}$PS$_3$  measured at the indicated temperature.} \label{fig1}
\end{figure}

\begin{figure}[htb]
\centerline{\includegraphics[width=1.05\textwidth]{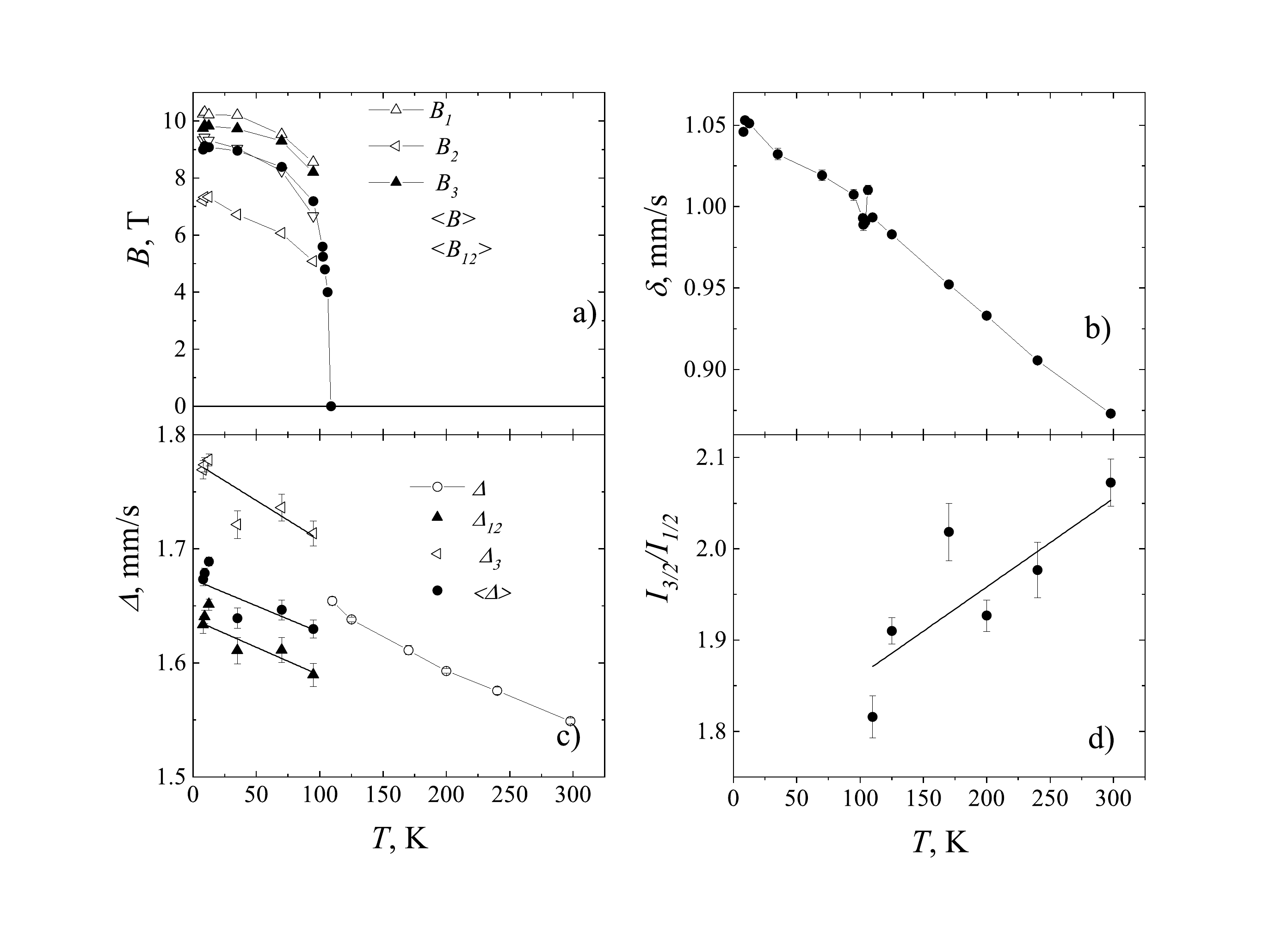}}
\caption{Dependence of hyperfine field $B$ (a), isomer shift $\delta$ (b), quadrupole splitting $\Delta$ (c) and intensity ratio of lines of doublet (d) of M\"{o}ssbauer spectra of Cu$_{0.15}$Fe$_{0.85}$PS$_3$  crystal on temperature.} \label{fig2}
\end{figure}

M\"{o}ssbauer spectra displays  the transition from an antiferromagnetic state to a paramagnetic state near the 110~K temperature (figure~\ref{fig1}). Above the antiferromagnetic transition temperature, M\"{o}ssbauer spectra were fitted to one doublet. On contrary, the magnetic subspectra (Hamiltonian method in Normos Site) should be used to fit the M\"{o}ssbauer spectra below the transition temperature. Previously, the spectra of FePS$_3$ were fitted using one set of hyperfine parameters (one subspectrum)~\cite{L4}. We found that for Cu$_{0.15}$Fe$_{0.85}$PS$_3$,  a good quality of fitting was achieved using three subspectra described by Hamiltonian method (WinNormos Site) in the case of single crystal~\cite{L5} except the spectra measured above 103~K temperature and below transition point which were broadened and needed extra subspectra to fit. These spectra were fitted using hyperfine field distribution. The set of Hamiltonian parameters included electric field gradient (EFG) asymmetry parameter $\eta$ and angles $\theta$ and $\beta$. $\theta$ is the angle between $z$ axis of EFG system and the direction of the magnetic field while $\beta$ is between $z$ axis of EFG system and $\gamma$-ray direction. The isomer shift $\delta$ and quadrupole coupling constant $eQV_{zz}/2$, where $Q$ is nuclear quadrupole moment, $V_{zz}$ is EFG tensor component, were the same for all three subspectra. The main area (70--75\%) of the M\"{o}ssbauer spectrum below 102~K can be attributed to two magnetically split subspectra having hyperfine fields $B_1$ and $B_2$ differing approximately by 1T (figure~\ref{fig2}~a). However, the effective hyperfine field of them $\langle B_{12}\rangle$ was very close to that found for FePS$_3$~\cite{L4, L6}. For these two subspectra, $\theta$ and $\beta$ angles were fixed to zero i.e., $\theta=0$ and $\beta=0$. In this case, $z$ axis of EFG and the magnetization were normal to \emph{ab}-plane similar to single-crystalline FePS$_3$. For both subspectra, the combined quadruple and magnetic interactions gave two sets of overlapping four lines (figure~\ref{fig1}). For the third subspectrum fitted to the experimental spectra, the angles were allowed to change and the best fits were obtained for $\theta\approx10$--$30^\circ$ and $\beta\approx20$--$40^\circ$. Moreover, the asymmetry parameter $\eta\approx0.76$ for the third subspectrum was larger than that of the first two subspectra ($\eta=0.23$--$0.28$). Therefore, taking the Cu atoms distribution into consideration, the third subspectrum can be considered to be due to distorted iron atom sites having Cu atoms as first neighbours in hexagonal arrangement of metal atoms along \emph{ab}-plane. Thus, the other two subspectra were attributed to iron sites similar to FePS$_3$~\cite{L4, L6}. It should be noted that FePS$_3$ undergoes first order phase transition at 120~K temperature. Coexistence of paramagnetic and antiferromagnetic phases was observed~\cite{L4}. For Cu$_{0.15}$Fe$_{0.85}$PS$_3$, the paramagnetic doublet starts also appear together with much more broadened magnetically split part of spectra above 102~K.

The lines of the doublet can be attributed to $\gamma$-ray transitions to $\pm3/2$ and $\pm1/2$ excited energy levels of $^{57}$Fe nucleus. The line intensity ratio of 1.8--2.03 found above the phase transition temperature (figure~\ref{fig2}~d) according to
$$
I_{3/4}/I_{1/2}=(1+\cos^2\beta)/(5/3-\cos^2\beta)
$$
gave the angle $\beta=27 $--$ 32^\circ$. Here, it is assumed that the EFG asymmetry parameter $\eta=0$. Quadruple splitting $\Delta=({eQV_{zz}}/{2})\left(1+{\eta^2}/{3}\right)^{1/2}$ was somewhat different below and above the antiferromagnetic transition point. This can be explained by the first-order transition because the lattice undergoes some transformations~\cite{L4}, as in the case of FePS$_3$ when $V_{zz}>0$.

\section{Magnetic moment measurements}

\subsection{Temperature dependence of the magnetic susceptibility}

\begin{figure}[htb]
	\centerline{\includegraphics[width=0.25\textwidth]{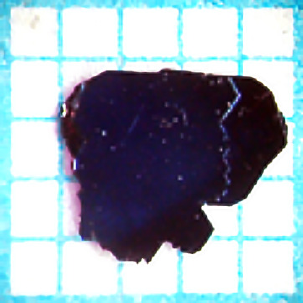}}
	\caption{(Colour online) The sample of  Cu$_{0.15}$Fe$_{0.85}$PS$_3$  crystal used for magnetic moment measurements.} \label{fig3}
\end{figure}

\begin{figure}[htb]
	\centerline{\includegraphics[width=0.75\textwidth]{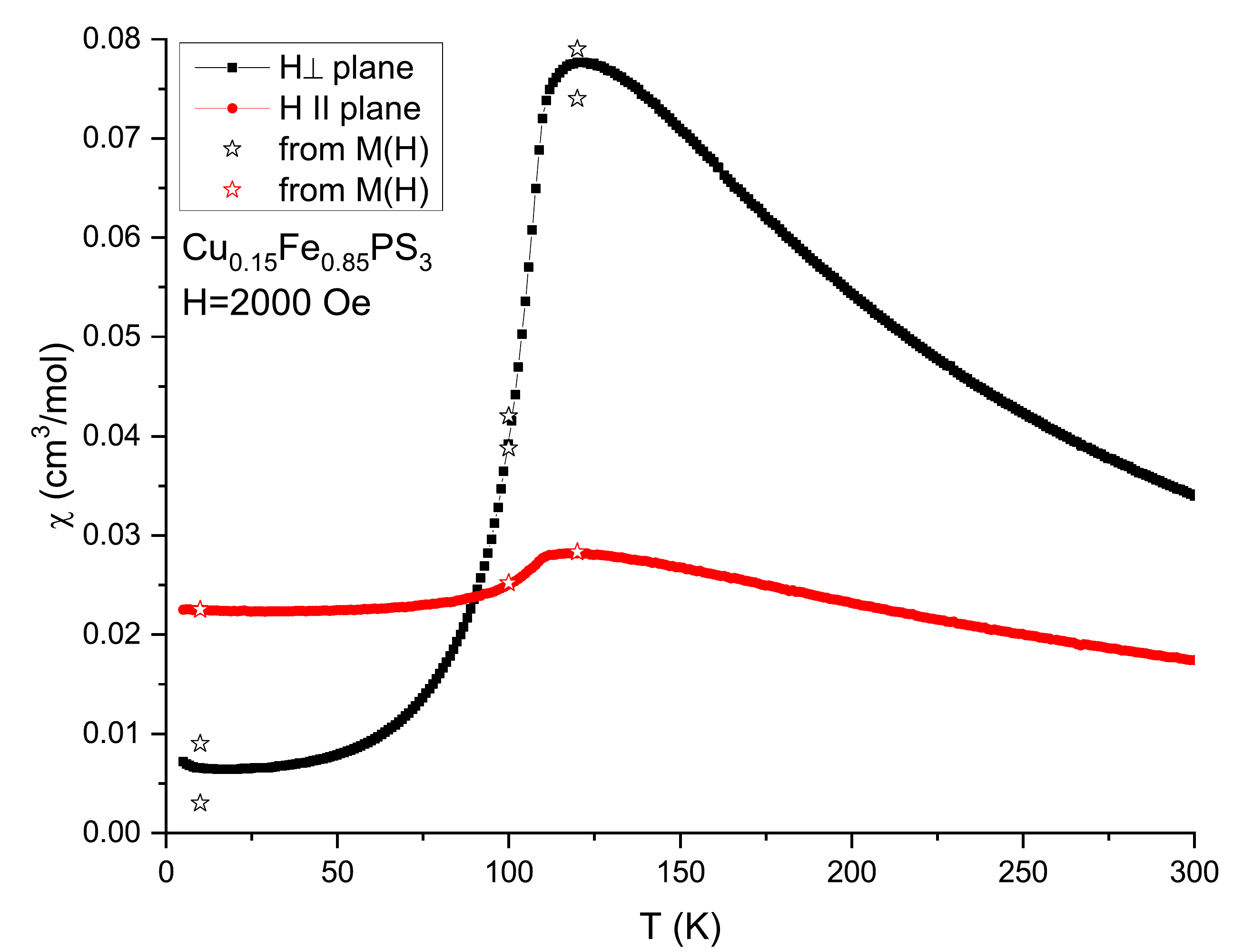}}
	\caption{(Colour online) Temperature dependence of the magnetic susceptibility $\chi(T)=M(T)/H$ of a Cu$_{0.15}$Fe$_{0.85}$PS$_3$  single crystal in magnetic field $H = 2000$~Oe for two directions $H$$\perp$ plane and $H$$\parallel$ plane of crystal layers.} \label{fig4}
\end{figure}

Single crystal Cu$_{0.15}$Fe$_{0.85}$PS$_3$ sample, of $\sim 3\times4$~mm flake-shaped (depicted in figure~\ref{fig3}) and 1.54~mg was used for magnetic moment measurements. For all orientations of the external magnetic field at a temperature of about 108~K, a sharp change in the temperature dependence of the magnetic susceptibility (see figure~\ref{fig4}) was observed, which may indicate the presence of a magnetic phase transition at this temperature.

No narrow anomaly associated with 3D ordering is observed on the $\chi(T)$ curve in the region of 108~K.
Therefore, it is possible that the ordering can also occur at a higher temperature, and in this case, we are dealing with a phase transformation of a magnetically ordered phase. For $H$$\perp$ plane of crystal layers and $T < 108$~K, as the temperature is lowered, the magnetic susceptibility $\chi(T)$ of the sample decreases significantly, which is typical when antiferromagnetic correlations are established in the spin system.

In addition to the experimental curves $\chi(T)$ (in a constant magnetic field $H=2000$~Oe) in figure~\ref{fig4}, the asterisks show several estimates of the values of the magnetic susceptibility of the Cu$_{0.15}$Fe$_{0.85}$PS$_3$ sample obtained by analyzing the linear sections of the field dependence $M(H)$ measured up to 5~T [see~$M(H)$ in the following figures]. As seen in figure~\ref{fig4}, the match for the orientation of the $H$$\parallel$ plane is perfect, with little deviation found between different experimental techniques for the $H$$\perp$ plane. The latter seems to be connected with the discovery of remanent magnetization (or a spontaneous weakly ferromagnetic moment in an antiferromagnet as a result of a small sublattice canting) in experiments for the $H$$\perp$ plane.

\subsection{Field dependence of the magnetic moment}

\begin{figure}[h!]
\centerline{\includegraphics[width=0.75\textwidth]{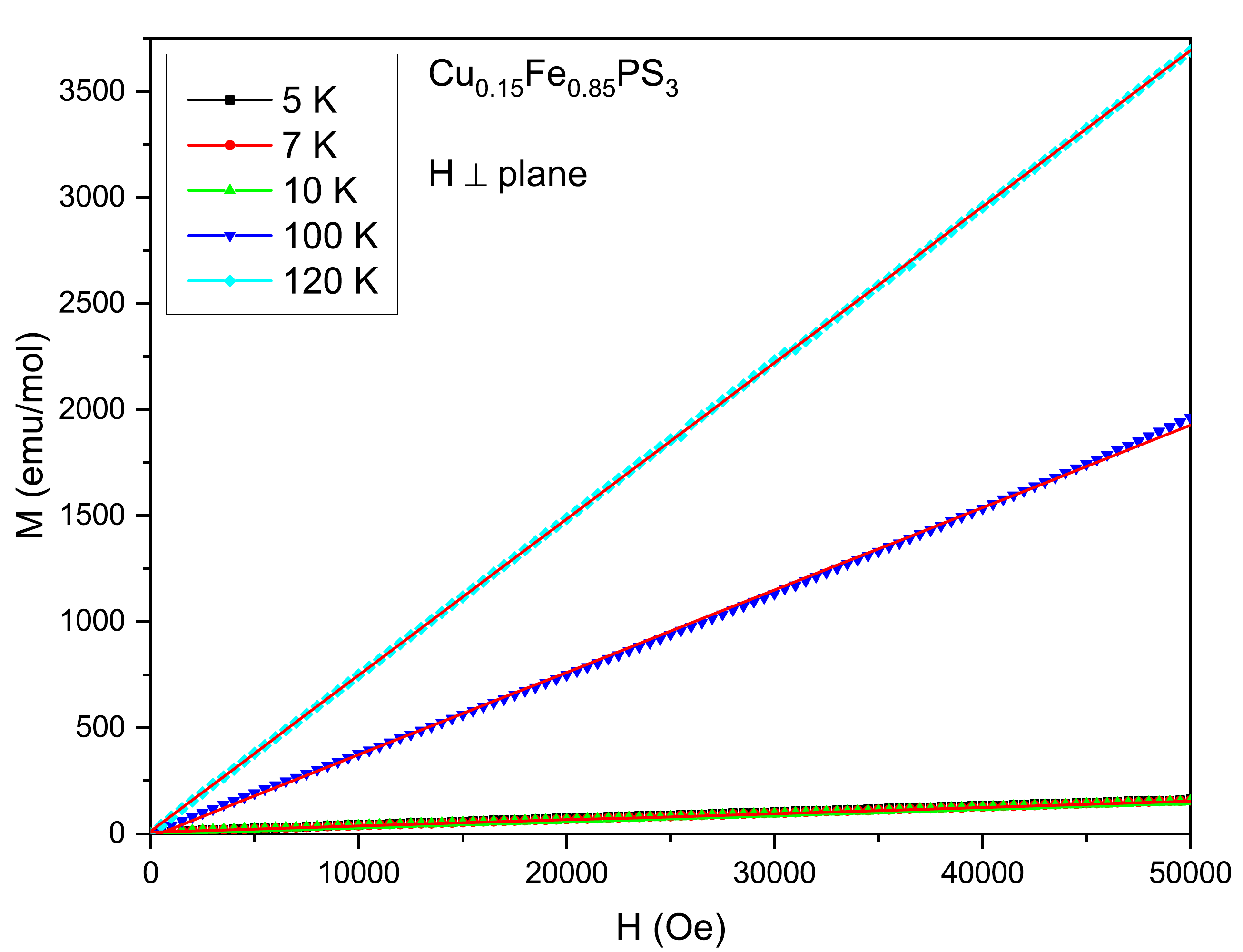}}
\centerline{\includegraphics[width=0.75\textwidth]{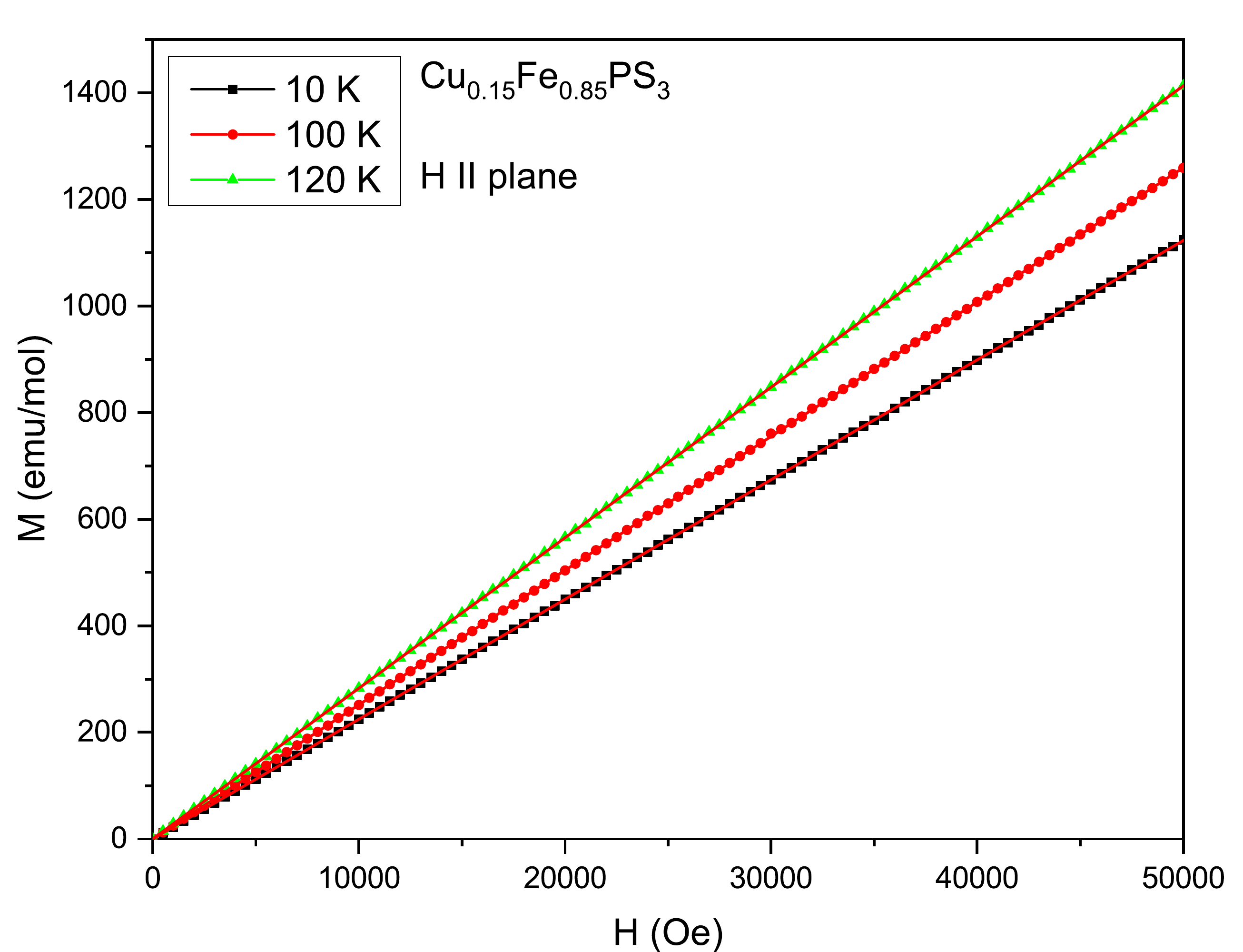}}
\caption{(Colour online) Field dependence of the magnetic moment  $M(H)$ of Cu$_{0.15}$Fe$_{0.85}$PS$_3$ sample at different temperatures for two orientations of the magnetic field:  $H$$\perp$ plane (top panel) and $H$$\parallel$ plane of crystal layers (bottom panel).} \label{fig5}
\end{figure}

The field dependencies of the magnetic moment of the Cu$_{0.15}$Fe$_{0.85}$PS$_3$ sample were studied at 120~K --- above the assumed phase transition, 100~K --- approximately the middle of the phase transformation, and 10~K --- the low-temperature point, where all the magnetic transformation processes have already been completed based on their general form $\chi(T)$. All the obtained curves $M(H)$ for the $H$$\perp$ plane and $H$$\parallel$ plane of crystal layers are shown in figure~\ref{fig5}. All of them demonstrate a fairly good linear behavior of $M(H)$ at any temperatures both above and below the assumed phase transition.

\begin{figure}[htb]
\centerline{\includegraphics[width=0.75\textwidth]{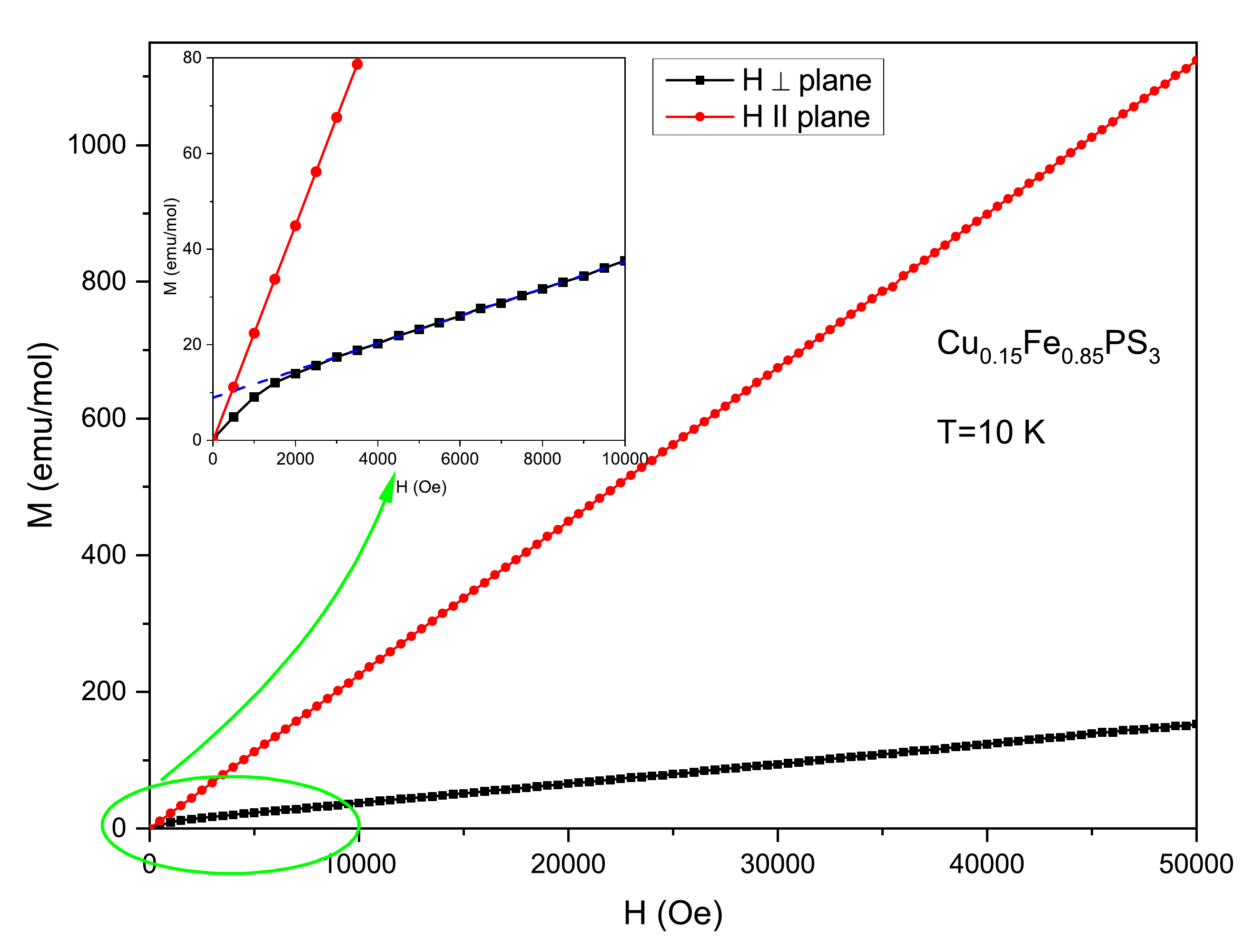}}
\caption{(Colour online) Field dependence of the magnetic moment of the Cu$_{0.15}$Fe$_{0.85}$PS$_3$ sample at a temperature of $T=10$~K for  $H$$\perp$ plane and $H$$\parallel$ plane. The insertion shows a low-field region.} \label{fig6}
\end{figure}

For the $H$$\perp$ plane at $T = 10$~K, extrapolation of the linear dependence (region above 2000 Oe) to zero magnetic fields detects the presence of a non-zero residual magnetic moment of the sample of the order of 9 emu/mol (dotted line in the insertion in figure~\ref{fig6}). This may be due to the existence of a canted magnetic sublattice in an ordered antiferromagnetic system and, hence, to a weak ferromagnetic moment of the system in this direction. At the same time, such a feature in $M(H)$ is completely absent for the other direction of the $H$$\parallel$ plane. This behavior of the magnetization for the $H$$\perp$ plane can be satisfactorily explained within the framework of the four-sublattice model of an antiferromagnet of the flat cross-type. In very weak fields, the ground state will be similar to the state with latent ferromagnetism and the total moment of the system will be close to zero. A rapid increase in the magnetization with the increasing field will occur due to the orientational flip of a skewed pair with a weakly ferromagnetic moment oriented against the field to the direction of the skewed pair in which this moment is oriented along the field. In the fields above 2000 Oe, the four-sublattice system effectively turns into a simple two-sublattice system, in which the brace between the sublattices gives a weak moment along the field.

Thus, a characteristic field of about 2000 Oe can be considered as a phase transition field between the effective four-sublattice and two-sublattice magnetic structures for the system under study.

\section{\emph{Ab initio} simulation results}

\begin{figure}[htb]
\centerline{\includegraphics[width=0.65\textwidth]{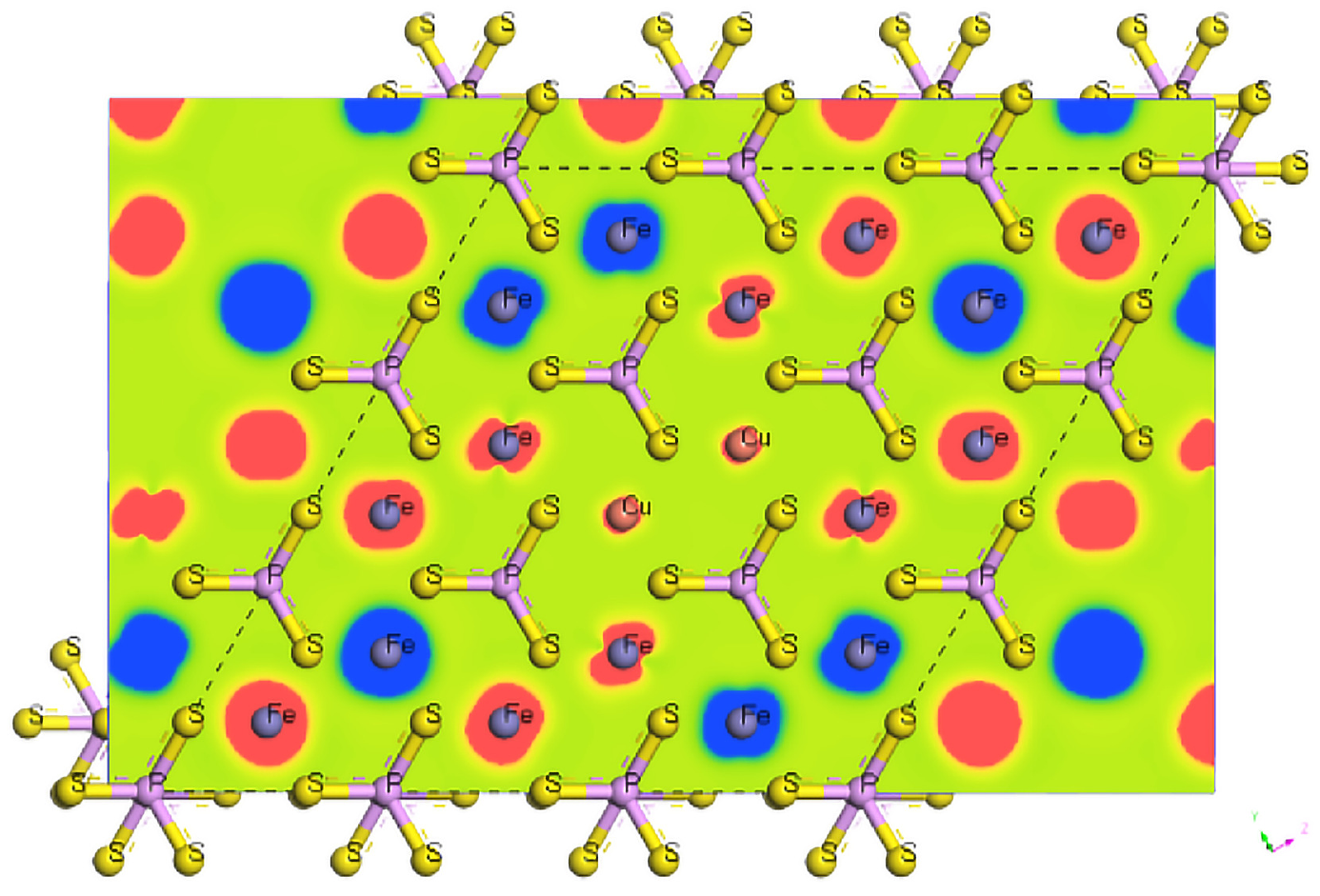}}
\caption{(Colour online) Spatial distribution of the calculated spin density of the Cu$_{0.15}$Fe$_{0.85}$PS$_3$ model at 0~K. Red and blue regions correspond to opposite spin orientations.} \label{fig7}
\end{figure}

The QUANTUM ESPRESSO package~\cite{L2}, was used to perform calculations. They were carried out within the generalized gradient approximation (GGA) with the Perdew--Burke--Ernzerhof functional~\cite{L7} or with the CA-PZ local functional based on the Ceperley and Alder data~\cite{L8} parameterized by Perdew and Zunger~\cite{L9}. The layered Cu$_{0.15}$Fe$_{0.85}$PS$_3$ crystal possesses the so-called vdW gap between layers. Therefore, the DFT-D method taking into account the dispersion interaction elaborated by Grimme~\cite{L10} is needed to calculate electronic properties of the material. The ultra-soft pseudopotential~\cite{L11} was used to perform calculations for Fe --- $3d^6$ $4s^2$, Cu --- $3d^{10}$ $4s^1$, P --- $3s^2$ $3p^3$, S --- $3s^2$ $3p^4$ atomic configurations. The plane-wave basis set cut-off was chosen to be equal to 600~eV. The Monkhorst--Pack $k$-points grid~\cite{L12} sampling was set at $12\times12\times3$ points for the Brillouin zone. The convergence tolerance parameters were as follows: energy $5\cdot10^{-6}$~eV, force 0.01~eV~\AA$^{-1}$; stress 0.02~GPa; displacement 0.05~\AA. The total energy convergence criterion was assumed to be fulfilled when the self-consistent field tolerance reaches the value $10^{-7}$~eV per atom. The \emph{ab initio} simulation of the features of electronic and spin subsystems in the framework of electron density functional theory (DFT) allows us to estimate the values of spin density (figure~\ref{fig7}) in the vicinity of all species' sites present in the investigated system. The supercell of $3\times3\times1$ unit cells of FePS$_3$ single-crystal was used for modelling the proper concentration of Cu substitutional atoms. A pair of Cu was used to preserve the symmetry of the Cu$_{0.15}$Fe$_{0.85}$PS$_3$ lattice.

Components of the EFG tensor and asymmetry parameter $\eta$ for Fe ions close to the experimental value for FePS$_3$ crystal ($\eta\approx0.23$--$0.28$) were calculated for the system under study at $T=0$~K (table~\ref{tbl1}).

\begin{table}[htb]
\caption{Calculated averaged components of the EFG tensor and asymmetry parameter for Cu$_{0.15}$Fe$_{0.85}$PS$_3$ crystal.} \label{tbl1}
\vspace{2ex}
\begin{center}
\renewcommand{\arraystretch}{0}
\begin{tabular}{|c||c|c|}
\hline
 Species & $\langle C_q \rangle$, MHz & $\langle \eta \rangle$ \strut\\
\hline
\rule{0pt}{2pt}&&\\
\hline
Cu & $-3.1501$ & 0.8797 \strut\\
\hline
Fe & 11.2971 & 0.2911  \strut\\
\hline
P  & 32.7594 & 0.2774  \strut\\
\hline
S  & $-26.1410$ & 0.0941  \strut\\
\hline
\end{tabular}
\renewcommand{\arraystretch}{1}
\end{center}
\end{table}

Furthermore, analysis of the Mulliken population shows that the main contribution to the ferromagnetic spin density is originated from $3d$-copper and $3p$-sulfur orbitals (see table~\ref{tbl2}).

\begin{table}[htb]
\caption{Decomposition of spin density (spin up -- spin down) contributions over orbitals in Cu$_{0.15}$Fe$_{0.85}$PS$_3$ crystal. The largest contributions are typed in bold.} \label{tbl2}
\vspace{2ex}
\begin{center}
\renewcommand{\arraystretch}{0}
\begin{tabular}{|c||c|c|c|c||c|}
\hline
Species & $3d$ & $3p$  & $4f$  & $4s$ & Total   \strut\\
\hline
\rule{0pt}{2pt}&&&&&\\
\hline
Cu & \bf $-0.248$ & $-0.028$ & 0.000 & $-0.002$ & $-0.278$ \strut\\
\hline
Fe &     $-0.087$ &  0.030 & 0.010 & $-0.002$ & $-0.049$ \strut\\
\hline
P  &     $-0.028$ & $-0.030$ & 0.000 &  0.000 & $-0.058$ \strut\\
\hline
S  &     $-0.072$ & \bf $-0.282$ & 0.000 & $-0.016$ & $-0.370$ \strut\\
\hline
\rule{0pt}{2pt}&&&&&\\
\hline
Total  & $-0.435$ &  $-0.310$ & 0.010 & $-0.020$ & $-0.755$ \strut\\
\hline
\end{tabular}
\renewcommand{\arraystretch}{1}
\end{center}
\end{table}

The values of the valence charge transfer and the spin polarization were estimated for the Cu$_{0.15}$Fe$_{0.85}$PS$_3$  crystal (table~\ref{tbl3}). The presence of uncompensated spin ordering is found.

\begin{table}[htb]
\caption{The calculated values of the spin components of the valence charge and the spin polarization in Cu$_{0.15}$Fe$_{0.85}$PS$_3$ crystal.} \label{tbl3}
\vspace{2ex}
\begin{center}
\renewcommand{\arraystretch}{0}
\begin{tabular}{|c||c|c|c|c|}
\hline
Species & $Q_{\rm {up}}$, e & $Q_{\rm {dw}}$, e & $Q$, e & $\Sigma S$, $\hbar/2$  \strut\\
\hline
\rule{0pt}{2pt}&&&&\\
\hline
Cu & $15.31 \pm 0.00$  & $15.45 \pm 0.00$ & $-3.53$  & $-0.28$ \strut\\
\hline
Fe &  $13.86 \pm 1.36$ & $13.86 \pm 1.37$ & $-27.51$ & $-0.07$ \strut\\
\hline
P  &  $7.16 \pm 0.03$  & $7.16 \pm 0.03$  &  12.22 & $-0.07$ \strut\\
\hline
S  &  $7.82 \pm 0.08$  & $7.83 \pm 0.07$  &  18.82 & $-0.35$ \strut\\
\hline
\end{tabular}
\renewcommand{\arraystretch}{1}
\end{center}
\end{table}

\section{Discussions}

In contrast to the expectation that Cu with spin $\frac12$ will dilute the magnetic moments contributed by Fe with a larger spin, we found that 15\% Cu doping partially keeps the effective fluctuating moment, although there is a long-range magnetic order partially distorted. This follows from the magnetic susceptibility temperature dependence around transition from paramagnetic into antiferromagnetic phase --- amplitude of the observed  $\chi(T)$ anomaly is like the one observed in case of FePS$_3$ pure crystal~\cite{L4}. At Fe by Cu partial substitution, the temperature of $\chi(T)$ maximum is lowered from 120~K in FePS$_3$ to 108~K in case of Cu$_{0.15}$Fe$_{0.85}$PS$_3$. The jump of $\chi(T)$ at ferromagnetic transition is smeared by Cu dilution.

In real crystal, the sulfur vacancies $V_{\rm S}$ presence can also effectively suppress the strong intralayer antiferromagnetic correlation, giving rise to a weak ferromagnetic ground state, which is observed on the $M(H)$ dependence below 10~K. The presence of $V_{\rm S}$ disrupts anion-mediated AFM interactions and may be responsible for the suppression of long-range AFM correlations. Herein, the  competing ferromagnetic exchange interactions can dominate at low temperatures, creating a magnetically frustrated system. The exchange interactions between the $V_{\rm S}$ and metal ions, and with the local atomic distortion in the vicinity of defects, could also induce ferromagnetism.

A canting configuration and the resulting net moments along the easy axis can also be attributed to atomic substitution. All these signatures exhibit the complexity of magnetic structure for the 15\% Cu substituted sample.

The susceptibilities of Cu$_{0.15}$Fe$_{1.85}$PS$_3$ can be considered as for configurationally averaged clusters sum of two terms: a randomly diluted antiferromagnet susceptibility, as in pure FePS$_3$, and a Curie correction arising from local fluctuations of the uncompensated spins due to the finite size of the cluster. For Cu$_{0.15}$Fe$_{1.85}$PS$_3$ crystal, there is observed a weak paramagnetic contribution for small temperatures (figure~\ref{fig4}), because many of the spins no longer belong to the infinite cluster. The uncompensated moments in the diluted 2D AFM can give rise to a Curie contribution into the magnetic susceptibility, and the presence of field dependence of the magnetization suggests that they interact ferromagnetically to give a spontaneous magnetization below 10~K.

\section{Conclusions}

According to the presented experimental data on M\"{o}ssbauer spectroscopy and direct magnetic moment temperature and field dependencies measurements, it can be stated that the Cu$_{0.15}$Fe$_{0.85}$PS$_3$ single-crystal undergoes a magnetic phase transition at the region of $102$--$108$~K. A weak ferromagnetic moment at the low-temperature region ($T=10$~K) was observed in the $H$$\perp$ plane direction. This correlates with \emph{ab initio} calculated non-zero spin polarization of the considered material. Furthermore, the calculated values of the electric field gradient components and estimations of the total magnetic moment of the unit cell (0.764~$\hbar$/2 corresponds to 8.543 emu/mol) are in reasonable agreement with the measured experimental quantities of $\sim 9$~emu/mol.

The present studies of Cu$_{0.15}$Fe$_{0.85}$PS$_3$ magnetic properties show that the uncompensated moments in the diluted 2D AFM can give rise to a Curie contribution, and the observed field dependence of the magnetization suggests that they do interact ferromagnetically to give a spontaneous magnetization at low temperatures. The relation of the magnetic ordering with structural change at antiferromagnetic phase transition can also be important and must be investigated further on.

It is worth to note that the flattening of magnetic susceptibility at low temperature was also observed in rare-earth containing compounds caused by Kondo effect~\cite{LL21}. In the case of the studied Cu$_{0.15}$Fe$_{0.85}$PS$_3$ crystal, the system seems not to be such a strongly correlated material as rare earth materials but, of course, the proper investigation of the low temperature dependence of resistivity is needed.

%\bibliographystyle{cmpj}
%\bibliography{cmpj_bibliography}

\begin{thebibliography}{99}
	
%1
\bibitem{LL1} Chu J., Wang Y., Wang X., Hu K., Rao G., Gong C., Wu C., Hong H., Wang X., Liu K., Gao C., Xiong J., Adv.~Mater., 2020, \textbf{33}, 2004469, \doi{10.1002/adma.202004469}.
%2
\bibitem{LL2} Hu T., Kan E., WIREs Comput. Mol. Sci., 2019, \textbf{9}, e1409, \doi{10.1002/wcms.1409}.	
%3
\bibitem{LL3} Liu Z., Deng L., Peng B., Nano Res., 2021, \textbf{14}, 1802--1813, \doi{10.1007/s12274-020-2860-3}.	
%4
\bibitem{LL4} Gong C., Kim E. M., Wang Y., Lee G., Zhang X., Nat. Commun., 2019, \textbf{10}, 2657, \doi{10.1038/s41467-019-10693-0}.
%5
\bibitem{LL5} Wang F., Shifa T. A., Yu P., He P., Liu Y., Wang F., Wang Z., Zhan X., Lou X., Xia F., He J., Adv. Funct. Mater., 2018, \textbf{28}, 1802151, \doi{10.1002/adfm.201802151}.	
%6
\bibitem{LL6} Gong C., Li L., Li Z., Ji H., Stern A., Xia Y., Cao T., Bao W., Wang C., Wang Y., Qiu Z. Q., Cava~R.~J., Louie~S.~G., Xia~J., Zhang X., Nature, 2017, \textbf{546}, 265--269, \doi{10.1038/nature22060}.	
%7
\bibitem{LL7} Belianinov A., He Q., Dziaugys A., Maksymovych P., Eliseev E., Borisevich A., Morozovska A., Banys J., Vysochanskii Yu., Kalinin S. V., Nano Lett., 2015, \textbf{15}, 3808--3814, \doi{10.1021/acs.nanolett.5b00491}.
%8
\bibitem{LL8} Liu F., You L., Seyler K. L., Li X., Yu P., Lin J., Wang X., Zhou J., Wang H., He H., Pantelides S. T., Zhou W., Sharma P., Xu X., Ajayan P. M., Wang J., Liu Z., Nat. Commun., 2016, \textbf{7}, 12357, \doi{10.1038/ncomms12357}.
%9
\bibitem{LL9} Qi J., Wang H., Chen X., Qian X., Appl. Phys. Lett., 2018, \textbf{113}, 043102, \doi{10.1063/1.5038037}.
%10
\bibitem{LL10} Osada M., Sasaki T., APL Mater., 2019, \textbf{7}, 120902, \doi{10.1063/1.5129447}.
%11
\bibitem{LL11} Lai Y., Song Z.,  Wan Y., Xue M., Wang C., Ye Y., Dai L., Zhang Z., Yang W., Dua H., Yangace J., Nanoscale, 2019, \textbf{11}, 5163--5170, \doi{10.1039/C9NR00738E}.
%12
\bibitem{LL12} Dziaugys A., Shvartsman V. V., Macutkevic J., Banys J., Vysochanskii Yu., Kleemann W., Phys. Rev. B, 2012, \textbf{85}, 134105, \doi{10.1103/PhysRevB.85.134105}.
%13
\bibitem{LL13} Kleemann W., Shvartsman V. V., Borisov P., Banys J., Vysochanskii Yu. M., Phys. Rev. B, 2011, \textbf{84}, 094411,\\ \doi{10.1103/PhysRevB.84.094411}.
%14
\bibitem{LL14} Olsen T., J. Phys. D: Appl. Phys., 2021, \textbf{54}, 314001, \doi{10.1088/1361-6463/ac000e}.
%15
\bibitem{LL15} Wildes A. R., Zhitomirsky M. E., Ziman T., Lan\c{c}on D., Walker H. C., J. Phys. D: Appl. Phys., 2020, \textbf{127}, 223903, \doi{10.1063/5.0009114}.
%16
\bibitem{LL16} Sakai H., Yamazaki T., Machida N., Shigematsu T., Nasu S., Mol. Cryst. Liq. Cryst. Sci. Technol., Sect. A, 2000, \textbf{341}, 105--110, \doi{10.1080/10587250008026125}.
%17
\bibitem{LL17} Chandrasekharan N., Vasudevan S., Phys. Rev. B, 1996, \textbf{54}, 14903, \doi{10.1103/PhysRevB.54.14903}.
%18
\bibitem{LL18} Mulders A. M., Klaasse J. C. P., Goossens D. J., Chadwick J., Hicks T. J., J. Phys.: Condens. Matter, 2002, \textbf{14}, 8697, \doi{10.1088/0953-8984/14/37/306}.
%19
\bibitem{LL19} Masubuchi T., Hoya H., Watanabe T., Takahashi Y., Ban S., Ohkubo N., Takase K., Takano Y., J. Alloys Compd., 2008, \textbf{460}, 668--674, \doi{10.1016/j.jallcom.2007.06.063}.
%20
\bibitem{LL20} Bhutani A., Zuo J. L., McAuliffe R. D., dela Cruz C. R., Shoemaker D. P., Phys. Rev. Mater., 2020, \textbf{4}, 034411,\\ \doi{10.1103/PhysRevMaterials.4.034411}.
%21
\bibitem{L1} Dziaugys A., Macutkevic J., Svirskas S., Juskenas R., Wencka M., Banys J., Motria S. F., Vysochanskii Yu., J.~Alloys~Compd., 2015, \textbf{650}, 386--392, \doi{10.1016/j.jallcom.2015.07.261}.
%22
\bibitem{L3} WissEl, Data Evaluation Software --- WissEl, Wissenschaftliche Elektronik GmbH, [Online; accessed 10-Jul-2022], URL \url{http://www.wissel-gmbh.de/index.php?option=com_content&task=view&id=55&Itemid=116}.
%23
\bibitem{L4} Jernberg P., Bjarman S., W\"{a}ppling R., J. Magn. Magn. Mater., 1984, \textbf{46}, 178--190, \doi{10.1016/0304-8853(84)90355-X}.
%24
\bibitem{L5} Chen Y. L., Yang D. P., M\"{o}ssbauer Effect in Lattice Dynamics: Experimental Techniques and Applications, Wiley-VCH Verlag GmbH \& Co. KGaA, Weinheim, 2007.
%25
\bibitem{L6} Rule K. C., Cashion J. D., Mulders A. M., Hicks T. J., Hyperfine Interact., 2002, \textbf{141}, 219--222,\\ \doi{10.1023/A:1021286910897}.
%26
\bibitem{L2} Giannozzi P., Baroni S., Bonini N., Calandra M., Car R., Cavazzoni C., Ceresoli D., Chiarotti~G.~L., Cococcioni~M., Dabo~I. et al., J. Phys.: Condens. Matter, 2009, \textbf{21}, 395502, \doi{10.1088/0953-8984/21/39/395502}.
%27
\bibitem{L7} Perdew J. P., Burke K., Ernzerhof M., Phys. Rev. Lett., 1997, \textbf{78}, 1396, \doi{10.1103/PhysRevLett.78.1396}.
%28
\bibitem{L8} Ceperley D. M., Alder B. J., Phys. Rev. Lett., 1980, \textbf{45}, 566, \doi{10.1103/PhysRevLett.45.566}.
%29
\bibitem{L9} Perdew J. P., Zunger A., Phys. Rev. B, 1981, \textbf{23}, 5048, \doi{10.1103/PhysRevB.23.5048}.
%30
\bibitem{L10} Grimme S., J. Comput. Chem., 2006, \textbf{27}, 1787--1799, \doi{10.1002/jcc.20495}.
%31
\bibitem{L11} Vanderbilt D., Phys. Rev. B, 1990, \textbf{41}, 7892, \doi{10.1103/PhysRevB.41.7892}.
%32
\bibitem{L12} Monkhorst H. J., Pack J. D., Phys. Rev. B, 1976, \textbf{13}, 5188, \doi{10.1103/PhysRevB.13.5188}.
%33
\bibitem{LL21} Menth A., Buehler E., Geballe T. H., Phys. Rev. Lett., 1969, \textbf{22}, 295, \doi{10.1103/PhysRevLett.22.295}.
	
\end{thebibliography}

\newpage

\ukrainianpart

\title{Антиферомагнітний фазовий перехід в шаруватому напівпровіднику Cu$_{0.15}$Fe$_{0.85}$PS$_3$: експеримент та DFT моделювання}
\author{
        В.~Пащенко\refaddr{label1},
        О.~Блудов\refaddr{label1},
        Д.~Балтрунас\refaddr{label2},
        К.~Мазейка\refaddr{label2},
        С.~Мотря\refaddr{label3},
        К.~Глухов\refaddr{label3},
        Ю.~Височанський\refaddr{label3}}
\addresses{
\addr{label1} Інститут фізики і техніки низьких температур ім. Б. Вєркіна НАН України, просп. Науки 47, 61103, Харків, Україна
\addr{label2} Відділ ядерних досліджень Центру фізичних наук та технології, просп. Саваноріу 231, LT-02300, Вільнюс, Литва
\addr{label3} Інститут фізики і хімії твердого тіла, Ужгородський національний університет, вул. Волошина 54, 88000, Ужгород, Україна
}

\makeukrtitle

\begin{abstract}
\tolerance=3000%
Представлені експериментальні дослідження парамагнітно-антиферомагнітного фазового переходу методом мессбауерівської спектроскопії та вимірювання температурних і польових залежностей магнітної сприйнятливості в шаруватому кристалі Cu$_{0.15}$Fe$_{0.85}$PS$_3$. Особлива поведінка польової залежності намаг\-ніченості в області низьких температур свідчить про слабкий феромагнетизм досліджуваного матері\-алу. За допомогою \emph{ab initio} моделювання електронної та спінової підсистем, в рамках теорії функціоналу електрон\-ної густини, проаналізовано особливості спінового впорядкування при низькій температурі, а також зміни міжатомних взаємодій поблизу атомів заміщення Cu. Розраховані компоненти тензора градієнта електричного поля та параметра асиметрії для іонів Fe близькі до значень знайдених з мессбауерівських спектрів. Маллікенівські заселеності показують, що основний внесок у феромагнітну спінову густину вносять $3d$ орбіталі міді та $3p$ сірки. Розрахунковий загальний магнітний момент елементарної комірки (8.543~emu/mol) цілком узгоджується з виміряним експериментальним значенням $\sim9$~emu/mol.
\keywords метал-фосфорні трихалькогеніди, магнітне впорядкування, мессбауерівська спектроскопія, фазові переходи, теорія функціоналу електронної густини

\end{abstract}

\end{document}